\documentstyle[prd,aps,preprint]{revtex}
\tightenlines
\newcommand {\be}{\begin{eqnarray}}
\newcommand {\ee}{\end{eqnarray}}
\input epsf.tex
\def\DESepsf(#1 width #2){\epsfxsize=#2 \epsfbox{#1}}
\begin{document}

\draft
\preprint{\vbox{
\hbox{UMD-PP-00-51}
\hbox{SMU-HEP-00-01}
}}
\title{Mirror Dark Matter and Galaxy Core densities}
\author{ R. N. Mohapatra$^1$ and V. L. Teplitz$^{2, 3}$}

\address{$^1$Department of Physics, University of Maryland, College Park, MD, 20742\\
$^2$ Department of Physics, Southern Methodist University, Dallas, TX-75275\\
$^3$ Address until
06/30/2001, Office of Science and Technology Policy, Eexcutive Office of
the President, Washington, DC 20502.}
\date{January, 2000}
\maketitle
\begin{abstract}
We present a particle physics realization of a recent suggestion by
Spergel and Steinhardt
that  collisional but dissipationless dark matter may resolve the
core density problem in dark matter-dominated galaxies such as the dwarf
galaxies. The realization is the asymmetric mirror universe
model introduced to explain the neutrino puzzles and the microlensing
anomaly. The mirror baryons are the dark matter particles with the desired
properties. The time scales are right for resolution of
the  core density problem and formation of mirror stars (MACHOs observed
in microlensing experiments). The mass of the region homogenized by Silk
damping is between a dwarf and a large galaxy.
\end{abstract}

\section{Introduction}
Dark matter constitutes the bulk of the matter in the universe and a
proper understanding of the nature of the new particle that plays this
role has profound implications not only for cosmology but also for
particle physics beyond the standard model\cite{book}. It is therefore not
surprising that one of the major areas of research in both particle
physics and cosmology continues to be the physics of dark
matter.
Apart from the simple requirement that the right particle physics
candidate must have properties that yield the requisite relic
density and mass to dominate the mass content of the universe, it
should be required to
 provide a satisfactory resolution of three puzzles of dark matter
physics: (i) why is it that the contribution of baryons to the
mass density ($\Omega$) of the universe
is almost of the same order as the contribution of the dark matter to it ?
(ii) how does one understand the dark objects with mass $\sim 0.5
M_{\odot}$ observed in the MACHO experiment\cite{micro}, which are
supposed to constitute up to 50\% of the mass\cite{gyuk} of the halo of
the
Milky way galaxy and presumably be connected to the dark constituent that
contributes to $\Omega$ (in a manner that satisfies the ``environmental
impact'' conditions of Freese et al\cite{freese}? and, finally (iii) what
explains the density
profile of dark matter in galactic halos-- in particular, the apparent
evidence in favor of the fact that the core density of
galactic halos remain constant as the radius goes to zero.

There are many particle physics candidates for the dark constituent of
the universe. Generally speaking, the prime consideration that leads to
such candidates is that they yield the right order of magnitude for the
relic density and mass necessary to get the desired $\Omega_{DM} \approx
0.2-1$. This is, of course, the minimal criterion for any such candidate
and requires that the annihilation cross section of the
particles must be in a very specific range correlated with their mass. The
most widely discussed candidates are
the lightest supersymmetric particle (LSP) and the
Peccei-Quinn particle, the
axion. The first is expected to have a mass in the range of 100 GeV
whereas the mass of the second would be in the range of $\sim 10^{-6}$ eV.
Compare these values
with the proton mass  of one GeV). To understand within these models
why $\Omega_B \sim
\Omega_{DM}$, one needs to work in a special range for the parameters
of the theory. In either of these pictures, the MACHO observations must
have a separate explanation. Thus it may not be unfair to say that these
two most popular candidates do not resolve the first two of the three dark
matter puzzles.
In recent years it has been emphasized that the LSP and the axion may also
have
difficulty in explaining the observed core density behaviour of dwarf
spheroidal galaxies which are known to be dark matter dominated. The point
here is that both the axions and neutralinos, being collisionless and
nonrelativistic, accumulate at the core of galactic halos, leading to
a core density $\rho (R)$ which goes like $R^{-2}$ rather than a constant
which seems to fit data better\cite{primack,moore,frenk}. We will refer to
this as the core density puzzle.
This last puzzle has motivated Spergel and Steinhardt\cite{ss} to revive
and reinvigorate an old idea\cite{hall} that dark matter may be strongly
self interacting,
which for the right range of the parameters of the particle may
lead to a halo core which is much less dense and hence in better
agreement
with observations. To be more specific, it was noted in \cite{ss} that
if the dark matter particle is self interacting and has mean free
path of collision of about a kpc to a Mpc, then the core on this scale
cannot ``keep on accumulating'' dark matter particle, since these will now
scatter
and ``diffuse out''. For typical dark matter particle densities of order
of one particle per cm$^3$, this requires a cross section for scattering
of $\sigma\simeq 10^{-21}-10^{-24}$ cm$^2$. Furthermore, in order to
prevent dissipation which would lead to cooling and collapse to the core,
one has a lower limit on the mass of the exchanged particle that
must
exceed typical ``virial'' energy of particles ($\sim$ keV). An alternative
possibility is that the core is optically thick to exchanged particles. If
these
considerations stand the test of time, a theoretical challenge would be to
look for alternative dark matter candidates (different from the popular
ones described above) and the associated scenarios for physics beyond the
standard model.
A class of models known as mirror universe models have recently been
discussed. These are motivated theoretically by string theory and
experimentally by neutrino physics.
These models predict the existence of a mirror sector of
the universe with matter and force content identical to the familiar
sector (prior to symmetry
breaking)\cite{lee,berezhiani,volkas}. Symmetry breaking may either keep
the mirror symmetry exact or it may break it. This leads to two classes
of
mirror models: the symmetric mirror model, where all masses and forces in
the two sectors remain the same after symmetry breaking\cite{volkas} and
the asymmetric mirror model\cite{berezhiani} where the masses in the
mirror sector are larger than those in the familiar sector. The
mirror particles interact with the mirror photon and not the familiar
photon so
that they remain dark to our observations. Since the the lightest
particles of the mirror sector (other than the neutrinos), the mirror
proton and the mirror electron
(like in the familiar sector) are stable and will have abundances similar
to the
familiar protons and electrons, the proton being heavier could certainly
qualify as a dark matter candidate. It has indeed
been pointed out\cite{blin,tm} that, consistent with the cosmological
constraints
of the mirror universe theory, the mirror baryons have the desired relic
density  to play the role of dark matter of the universe.
 The additional neutrinos of the
mirror sector are the sterile neutrinos which appear to be needed in order
to have a simultaneous understanding of the three different neutrino
oscillations i.e. solar, atmospheric and the LSND observations.
In fact, one view of neutrino oscillation explanations of these phenomena
fixes the ratio of familiar particle mass to the mirror particle mass
thereby narrowing down the freedom of the mirror sector parameters. If
indeed sterile neutrinos turn out to be  required, mirror universe theory
is one of the few models where they appear naturally with mass in the
desired range. If
we denote the mass ratio $m_{p'}/m_p~=~\zeta$, then a value of $\zeta\sim
10$ is required to explain the neutrino puzzles.
What is more interesting is that for the
same range of parameters that are required to solve the neutrino puzzles,
(i.e. $\zeta \simeq 10$)
mirror matter can also provide an explanation of the
microlensing observations\cite{tm}- in particular  why the observed MACHOs
have a mass very near the solar mass and are still dark.

It is the goal of this paper to show that the same mirror universe
framework can also explain the core density puzzle of galactic halos.
The basic idea is that mirror sector $H'-H'$ (i.e. mirror hydrogen)
scattering, with its large geometric cross section, is a natural
candidate for strongly interacting dark matter of Ref.\cite{ss}. Thus the
mirror matter model has the desirable properties
that it can naturally explain all three dark matter puzzles. It is worth
noting that the asymmetric mirror model was not originally designed for
this purpose but rather to explain the neutrino puzzles and indeed it is
gratifying that slight modification of the framework (increasing the
QCD scale) that solves the neutrino puzzles also solves the dark matter
puzzles.

\section{Mirror matter as dark matter}

Let us start with a brief overview of the mirror matter models
\cite{berezhiani,volkas}. The basic idea of the model is extremely simple:
{\it  duplicate the standard model or any extension of it in the gauge
summetric Lagrangian (and allow for the possibility that symmetry
breaking may be different in the two sectors).}
There is an exact mirror symmetry connecting the Lagrangians (prior to
symmetry breaking)
describing physics in each sector. Clearly the $W's, \gamma's $ etc of
each sector are different from those in the other as are the quarks and
leptons. When the symmetry breaking scale is
different in the two sectors, we will call this the asymmetric mirror
model\cite{berezhiani}. The QCD scale being an independent scale in the
theory could be arbitrary. We will allow both
the weak scale as well as the QCD scale of the mirror sector to be
different from that of the familiar sector\cite{tm} and assume the same
common
ratio $\zeta$ for both scales i.e. $<H'>/<H>=\Lambda'/\Lambda\equiv
\zeta$.
It is assumed that the two sectors in the
universe are connected by only gravitational interactions. It was shown in
\cite{berezhiani,volkas} that gravity induced nonrenormalizable operators
generate mixings between the familiar and the mirror neutrinos, which is
one of the ingredients in the resolution of neutrino puzzles. It is of
course clear that both sectors of the universe are co-located.
Together they evolve according to
the rules of the usual big bang model except that the cosmic soups in
the two sectors may have different temperature. In fact the constraints of
big bang nucleosynthesis require that the post inflation reheat
temperature
in the mirror sector $T'_R$ be slightly lower than that in the familiar
sector $T_R$ (define $\beta\equiv T'_R/T_R$) so that the contribution of
the light mirror particles such as $\nu', \gamma'$ etc. to nucleosynthesis
is not too important. This has been called asymmetric inflation and can be
implemented in different ways\cite{asym}. Present discussions of BBN can
be used to conclude that roughly $\beta^4 \simeq 1/6$ is equivalent to
$\delta N_{\nu} \leq 1$.
Before proceeding to any detailed discussion, let us first note the impact
of the asymmetry on physical parameters and processes. First it implies
that $m_i\rightarrow\zeta\,m_i$ with $i=n,p,e,W,Z$. This has important
implications which have been summarized before\cite{berezhiani,teplitz}.
For instance, the binding energy of mirror hydrogen is $\zeta$ times
larger so that recombination in the mirror sector takes place
much earlier than in the visible sector. With $\beta\equiv T'_R/T_R$ as
above, mirror reombination
temperature is $\zeta/\beta T_r$ where $T_r$ is the recombination
temperature in the familiar sector. The mirror sector recombination takes
place before familiar sector recombination; this means that density
inhomogeneities in the mirror sector begin to grow earlier and familiar
matter can fall into it later.
 One can also compute the
contribution of mirror baryons to the mass density of the universe as
follows:
\begin{eqnarray}
\frac{\Omega_{B'}}{\Omega_{B}}\simeq \beta^3 \zeta
\label{cdm}
\end{eqnarray}
Here we have assumed that baryon to photon ratio in the familiar and the
mirror sectors are the same as would be expected since the dynamics are
same in both sectors due to mirror symmetry.
Eq. (1) implies that both the baryonic and the mirror baryon contribution
to $\Omega$ are roughly of the same order, as observed. This provides a
resolution of the first conceptual puzzle. Furthermore if
we take $\Omega_B\simeq 0.05$, then $\Omega_{B'}\simeq 0.2$ would require
that $\beta =(4/\zeta)^{1/3}$. From this one can calculate the effective
$\delta N_{\nu}$ using the following formula:
\begin{eqnarray}
\delta N_{\nu}= 3 \beta^4 + \frac{4}{7} \beta^4 (\frac{11}{4})^{4/3}
\end{eqnarray}
where the last factor $(11/4)^{4/3}$ is due to the reheating of the mirror
photon gas subsequent to mirror $e^{+'}e^{-'}$ annihilation. For $\zeta =
20$, this implies $\delta N_{\nu} \simeq 0.6$ and it scales with $\zeta$
as $\zeta^{-4/3}$. Thus in principle the idea
that mirror baryons are dark matter could be tested by more accurate
measurements of primordial He$^4$, Deuterium and Li$^7$ abundances which
determine $\delta N_{\nu}$.
Clearly to satisfy the inflationary constraint of $\Omega_{TOT}=1$, we
need
$\Omega_{\Lambda}\simeq 0.7$. These kinds of numbers for cold dark matter
density emerge from current type I supernovae
observations. It is interesting to note that if one were to require that
$\Omega_{CDM} = 1$, the mirror model would require that $\zeta$ be much
larger (more than 100) which would then create difficulties in
understanding both the neutrino data and the microlensing
anomalies. Thus mirror baryons satisfy the most requirement to
 be the cold dark matter of the universe.

\section{Mirror matter collision and core density of galaxies}
To discuss further implications of the mirror cold dark matter for
structure formation and the nature of the dark halos, we need to know
various cross sections. Using the asymmetry factor $\zeta$, it is easy to
see that weak cross sections varies as $\zeta^{-4}$ i.e. $\sigma'_W =
\sigma_W/\zeta^4$, the Thomson cross section $\sigma'_T=\sigma_T/\zeta^2$.
Nuclear cross sections will also be different in the mirror sector due to
different values of the QCD scale in the two sectors.
For $\Lambda' \simeq (10-15) \Lambda$,
 we would expect $\sigma'_{N'N'}= (\sigma_{NN}/\zeta^2)\times
\left(\frac{g_{\pi'N'N'}}{g_{\pi N N}}\right)^4$ (for fixed values of
energy). With these simple rules, assuming the pion nucleon couplings in
both sectors to be identical, we find that $\sigma_{N'N'}$
to be of order $10^{-30}$ cm$^2$ or so. This is clearly too small to
make a difference in the core density problem.
Let us now focus on the atomic forces. We would expect the mirror dark
matter to be mostly in the form of atomic hydrogen ($H'$). In the familiar
sector, the hydrogen atomic scattering is of order $\pi a^2_0$, (where
$a_0$ is the Bohr radius) and is of order $10^{-16}$ cm$^2$. Since the
Bohr radius
scales like $\zeta^{-1}$ when we consider the mirror sector, we would
expect the collisional mean free path due to atomic scattering to be of
order $10^{17}\zeta^3$ cm. For $\zeta =20$, the mean free path is about
$0.3$ kpc. This is slightly smaller than the lowest allowed value given in
Ref. \cite{ss}. However, as we argue below, the dangers for lower mean
free paths envisioned in Ref. \cite{ss}, do not apply to our case due to
the dynamics of the mirror universe.
As noted in Ref.\cite{ss}, an additional constraint on the self
interacting dark matter arises from the fact that the forces of self
interaction must not be such as to allow significant loss of energy
from  the core of halos since otherwise, core will lose energy and
collapse giving us back the problem we wanted to cure in the first place.
Essentially similar reasoning led to the lower bound of 1 kpc on the mean
free path in the Ref.\cite{ss}. In the mirror dark matter model, even
though we have mirror photons, there is no dissipation problem.
     The point is that when there is a collision, it will
excite the atom.  The atom radiates when it falls back down, but that
radiation can be absorbed.  Sometimes you ionize, then you get a plasma
which also absorbs the radiation.  The result would seem to be something
like the sun in which it takes a long time for the radiation to get out.
In other words, the mirror matter core is optically thick.
    In fact, one can estimate the number of collisions a core particle
makes since we know the mfp is in the range of 1 - 0.1 kpc. Taking
mfp/virial velocity, we get a time of 10-100Myr so it makes $10^3-10^4$
collisions during the age of the universe. We can also
give a crude estimate of the fractional energy loss per particle by
using $\sigma T^4$ for the energy radiated per
unit area per unit time, dividing by the number of particles and
multiplying by $10^{17}$ s - the age of the universe, and multiplying by
thearea of the ``core" of radius  1 kpc.  This fractional energy loss per
particle is negligible enabling us to have a lower mean free
path than the Ref.\cite{ss}. For the case of mirror matter,
the core is protected against this collapse by the long time it takes to
radiate away energy which is
in the form of (mirror) photons that can't get out. The essential point is
the similarity of the mirror sector with the familiar sector, where
we know that the photons emitted from the core of a star are very
few in number due to minimum ratio of surface to volume and therefore do
not lead to collapse of the galaxy cores.

\section{Structure Formation}
In this section we will try to make plausible that, in spite of the
$\zeta^2$ decrease in cross sections, the facts that (a) structure
formation begins earlier in the mirror sector (because recombination
occurs before matter-radiation equality) and consequently (b) mirror
temperatures are higher, for the same processes, than familiar
temperatures, permit formation of galactic and smaller structures.  In
doing this, we will make use of our previous work in \cite{teplitz} and
\cite{tm}, as well as that of Tegmark et al \cite{teg}.
Much of the work of \cite{teplitz} can be carried over to the present
work,
after suitable modification to take into account the fact that, in the
current model, the proton mass scales as $\zeta$.  Here, we will assume
that primordial perturbations are "curvature" or "adiabatic"
perturbations.  This means that the scale of the largest structures are
set by mirror Silk damping \cite{silk}.  $\gamma'$ diffusion wipes out
inhomogeneities until the $\gamma'$ mean free path,
\begin{equation}
                \lambda'=[\sigma_T\zeta^{-2}n_e^{'}]^{-1}
\end{equation}
where $\zeta^{-2}\sigma_T$ is the mirror matter Thomson cross section and
$n_e'$ is its electron number density, becomes greater than one third the
horizon distance ($ct$).  Silk damping turns off because the $\lambda'$
increases as $z^{-3}$ while $ct$ only increases as $z^{-2}$.
First, we compute, from Silk damping, the masses of the largest
structures in this picture.  Since structure formation starts with the
mirror sector, our assumption is that familiar sector particles will
later fall into these.  For numerical
values below, we will take, $h=0.7$ and $\Omega_{\tilde{B}}=0.2$.
We pick $t\sim(z_1/z)^2 s$ with $z_1=4\times10^9$ and
$n_{\tilde{e}}=\Omega_{tilde{B}}\rho_{c}z^3/(\zeta m_p)$ with
$\rho_c=1.9h^2\times10^{-29} g/cm^3$.
Silk damping stops at around $z_{sd}\sim 8\zeta^3$ which gives
\begin{eqnarray}
        \lambda_{sd}\sim2.5\times10^{27}/\zeta^6 ~~cm\\ \nonumber
        M_{sd}\sim10^{54}/\zeta^9~~gm
\end{eqnarray}
Note that, for $\zeta\sim 10$, this is about the mass (and size) of a
large
galaxy.  This coincidence could be an important factor in
understanding galaxy sizes should this model correspond to reality.
As in \cite{teplitz}, we parametrize the separation of $M_{sd}$ from the
expanding universe as taking place at
\begin{equation}
        z_{stop}=z_Mz_{sd}
\end{equation}
with
\begin{equation}
        R_G=\lambda_{sd}/z_M
\end{equation}
After violent relaxation we have for the proton temperature
\begin{equation}
        T_p=GM_G\zeta m_p/R_G \sim 10^{-4}z_M/\zeta^2 ~~ergs
\end{equation}
with $\rho$, outside the central plateau, given by
\begin{equation}
        \rho(R)=A/R^2 \sim 10^{26}z_M/(\zeta^3R^2)~~gm/cm^3
\end{equation}
We now turn to the question of whether this isothermal sphere is likely to
fragment and form mirror stars.  For this we compute the amount of mirror
molecular hydrogen since it is its collisional excitation (and
subsequent radiation) that is believed
to be the chief mechanism that provides cooling for formation of stars.
If the rate for this mechanism is faster than the rate for free fall into
the mass of the structure at issue, we can expect local regions to cool
fast enough to result in fragmentation of that
structure.  We do here a very rough estimate of mirror galaxy
fragmentation into mirror globular clusters, using the results of
\cite{teg}, but leave to a more detailed work further fragmentation into
the $0.5M_{\odot}$ structures predicted in \cite{tm}
Reference \cite{teg} give a useful approximation to their numerical
results for the fraction of molecular hydrogen, $f_2$ ($f_0$ denotes its
primordial value):
\begin{equation}
        f_2(t)=f_0+(k_m/k_1)ln[1+x_0nk_1t)
\end{equation}
where, as a first try, $k_m$ can be taken as just the rate for
$H+e^-\rightarrow H^-+\gamma$ (which they conveniently give as about
$2\times10^{-18}T^{.88} cm^3s^{-1}$), while $k_1$ is the rate for
$H^++e^-\rightarrow H+\gamma$ ($2\times10^{-10}T^{-0.64} cm^3s^{-1}$).
Equation
(8) is the result of $H_2$ production from the catalytic
reactions $H+e^-\rightarrow H^-+\gamma$ followed by $H+H^-\rightarrow
H_2+e^-$ competing
against the recombination reaction that destroys the catalyst, free
electrons,
(approximately) as $1/t$ (assuming constant density).  Our goal here is to
show from Equation (8) that it is plausible that $f_2$, the
fraction of molecular hydrogen, rises from its primordial value of
$10^{-6}$ to the region above $10^{-4}$ where cooling tends to be
competitive with free fall.
First, we note that, if $k=<v\sigma>\sim AT^{\gamma} cm^3/s$, for familiar
e and p, we expect that, for mirror e and p, scaling with $\zeta$ to go as
$\zeta^{-(2+\gamma)}AT^{\gamma}$, since $\sigma$ must go as $\zeta^{-2}$
and all factors of $T$ must be divided by some combination of $m_e$ and
$M_p$, both of which go as $\zeta$ (making this model much easier to
compute from than that of \cite{teplitz}).
We now estimate fragmentation.  From Equation (6) we see that the
galactic temperature should begin at about 10 eV at a time when the cosmic
temperature is about 1 eV and the cosmic gamma number density is about
$10^9/cm^3$.  The rate for ``compton cooling" is very fast at this high
density (unlike at later times for the familiar case) and there should be
rapid cooling to about 1 eV.  We can now compute the Jeans length for
fragments as a function of distance R from galaxy central.  We use
\begin{equation}
        \rho_J=(T/Gm)^3/M^2
\end{equation}
If we set the Jeans mass, $M$, to $4\pi r^3\rho_J/3$, we can solve for $r$
obtaining (if we are careful to convert $T$ in Equation (6)
from ergs)
\begin{equation}
        r=R[10^{-7}\zeta^2/z_M]^{1/2}\sim10^{-3.5}R
\end{equation}

Now inserting into Equation (8) gives the coefficient of the
log term on the order of $10^{-2}$ and the argument varying from
$10^{-13}$
to $10^{-17}$ as $R$ varies from 1 to 100 kiloparsecs while the free fall
time ($(G\rho)^{-1/2}$) varies from $10^{14}$ to $10^{16}$.  This would
appear to indicate the likelihood of fragmentation of the original Silk
damping structure into smaller units, (outside the optically thick core)
and the eventual formation of the
$0.5M_{odot}$ black holes that explain the microlensing events of
\cite{tm}.

\section{Conclusion}
The asymmetric mirror model\cite{berezhiani} was
originally proposed to solve the neutrino puzzles. Subsequently, it was
advocated\cite{tm,blin}  as providing an alternative dark matter
candidate. Then it was shown to have the advantage of resolving the
microlensing anomaly in a ``non-polluting'' manner\cite{freese}. In this
paper we have argued that the model could additionally provide an
explanation of a fourth problem. It appears to be a realization of the
mechanism of Spergel and steinhardt for understanding
 the core density profile of galaxies by means of
builtin self interactions of mirror matter.
 The
work of R. N. M. is supported by the National Science Foundation grant
under no. PHY-9802551 and the work of V. L. T. is supported by the DOE
under grant no. DE-FG03-95ER40908. We like to thank P. Steinhardt for
some discussions.

\end{document}